\documentclass[twocolumn]{aastex701}

\usepackage{marginnote}

\journalinfo{The Astrophysical Journal Supplement Series, 282:5 (15pp), 2026 January}

\begin{document}

\title{Measurements of Cosmic Proton Flux through Neutron Monitors Using Deep Networks and Imputation Techniques in the AMS-02 Era}

\author[orcid=0009-0009-1077-9226, sname='Zhao']{Pengwei Zhao}
\altaffiliation{These authors contributed equally to this work.}
\affiliation{School of Physics and Astronomy, Sun Yat-sen University, Zhuhai 519082, China}
\email{zhaopw5@mail2.sysu.edu.cn}  

\author[orcid=0000-0002-0491-8893, sname='Yan']{Jianqi Yan}
\altaffiliation{These authors contributed equally to this work.}
\affiliation{School of Science, Shenzhen Campus of Sun Yat-sen University, Shenzhen 518107, China}
\affiliation{Department of Physics, The University of Hong Kong, 999077, Hong Kong}
\email{yanjianqi.top@gmail.com}

\author[orcid=0000-0002-5652-6357, sname='Leung']{Alex P. Leung}
\affiliation{Department of Physics, The University of Hong Kong, 999077, Hong Kong}
\email{alexpl@hku.hk}

\author[orcid=0000-0003-3649-2476, sname='Feng']{Jie Feng}
\altaffiliation{Corresponding author}
\affiliation{School of Science, Shenzhen Campus of Sun Yat-sen University, Shenzhen 518107, China}
\email[show]{fengj77@mail.sysu.edu.cn}


\begin{abstract}

Accurate measurements of cosmic proton flux are essential for studying the modulation processes of cosmic rays during the solar activity cycle.
A proton flux measurement method, based on ground-based neutron monitor (NM) data and deep learning techniques, is presented.
After the necessary pre-processing of ground-based NM data using a convolutional neural network (CNN) model, we model the relationship between NM observations and proton flux measured by the Alpha Magnetic Spectrometer (AMS). 
The daily cosmic proton flux, ranging from 1 GV to 100 GV, is obtained for the period from 2011 to 2024, showing strong agreement with the observed values.
In addition, daily proton flux is computed for periods when AMS measurements were unavailable due to operational reasons. 
For the first time, hourly proton flux as a function of rigidity are calculated for the study of the short-time solar activities.

\end{abstract}

\keywords{\uat{Cosmic rays}{329} --- \uat{Forbush effect}{546} ---   \uat{Convolutional neural networks}{1938} --- \uat{Time series analysis}{1916}}


\section{Introduction}
\label{sec:level1}

Cosmic rays entering the solar system are modulated by the heliospheric magnetic field, whose strength and structure vary with solar activity. These variations influence and regulate the intensity of cosmic rays reaching the Earth's environment, producing observable fluctuations that closely correlate with solar activity levels.

The intensity of galactic cosmic rays is inversely correlated with solar activity. During periods of elevated solar activity, characterized by an increase in sunspot numbers, the cosmic-ray intensity correspondingly decreases. This phenomenon primarily follows an approximately 11-year sunspot cycle. Additionally, cosmic rays exhibit periodic variations on shorter timescales. For instance,~\citet{1939RvMP...11..173M} presented evidence of a 27-day modulation cycle, likely related to the solar rotation period.

In addition to these periodic variations, non-recurrent perturbations are also observed that are associated with sudden solar flares, which result in a rapid change in flux over a period ranging from several hours to days. Forbush Decreases (FDs) \citep{1971SSRv...12..658L} are examples of such perturbations, representing sudden decreases in Galactic Cosmic Rays (GCRs) due to intense solar wind transients.

The above phenomena can be observed by ground-based detectors (e.g., NMs \citep{2010JGRA..11512109M}) and space-based detectors, such as the Payload for Antimatter Matter Exploration and Light-nuclei Astrophysics (PAMELA) \citep{Adriani:2016uhu}, the Alpha Magnetic Spectrometer (AMS) \citep{AGUILAR20211}, and the Dark Matter Particle Explorer (DAMPE) \citep{2021ApJ...920L..43A}, which measure the time variation of cosmic rays.


Among them, NMs are a key type of ground-based detectors that provide long-term cosmic ray data for studies. After applying corrections for terrestrial factors, such as geomagnetic, atmospheric, and instrumental effects \citep{2021JGRA..12628941V}, these monitors detect secondary nucleons produced in the atmosphere from cascades initiated by primary cosmic-ray particles.


Since the Earth is shielded from the high-energy charged particles by the Geomagnetic field, the primary cosmic ray that interacts with the atmosphere should be with a rigidity greater than the Geomagnetic cutoff rigidity at the location of the NM station.

Since the establishment of the Climax monitor in 1951, NMs have served as essential instruments for cosmic ray observation. The global deployment of these monitors has progressively expanded \citep{2000SSRv...93..285M}, with approximately 50 stations currently operational within the international network. Each station records cosmic ray particles whose rigidity exceeds the local geomagnetic cutoff threshold. Consequently, the cosmic ray spectra detected at different monitoring locations encompass distinct rigidity ranges. Comprehensive technical specifications for all stations, including their respective cutoff rigidity values, are documented through the neutron monitor database network \citep{MAVROMICHALAKI20112210}.

Despite their extensive use in monitoring cosmic ray variations, NMs have certain limitations. Firstly, they measure the integrated flux of cosmic rays above the local geomagnetic cutoff rigidity (i.e., momentum per unit charge), without distinguishing between different particle species or their individual rigidities. This integration results in a combined measurement that encompasses all cosmic ray particles exceeding the cutoff threshold, therefore making it challenging to analyze specific contributions from various species or energy levels. Moreover, the cutoff rigidities of NMs do not reflect the actual rigidity of the cosmic ray flux, as low-energy cosmic rays do not produce enough secondary particles to reach the ground. 
\cite{2020JGRA..12527433M} calculated yield functions of NMs to connect NM measurements with cosmic ray flux. With the yield functions, \cite{2025JGRA..13033805V} simulated cosmic ray flux with NM counts and obtained good results between 5-15 GV. These traditional methods are indirect and model-dependent, so the uncertainties are large.



Unlike ground-based detectors, space-based detectors, such as AMS, are capable of directly detecting cosmic rays.
\cite{2021PhRvL.127A1102A} has provided detailed measurements of cosmic proton flux $\Phi_{p}$ between 2011 and 2019. The rigidity range spans from 1 to 100 GV and reveals periodic variations correlated with solar activity. In particular, the study highlights periodic flux variations on timescales of 27, 13.5, and 9 days, which are associated with solar rotation and the dynamics of interplanetary magnetic fields. 
In addition, \cite{PhysRevLett.128.231102} has measured the cosmic ray helium flux $\Phi_{\text{He}}$ during the same period.
The flux ratios $\Phi_{\text{He}}/\Phi_{p}$ are around 10$\%$ and vary with $\Phi_{p}$, suggesting that the NM count rates are likely correlated with $\Phi_{p}$.

While the AMS Collaboration has provided valuable insights into cosmic proton flux variations, challenges such as data discontinuities caused by detector studies and upgrades from September 2014 to November 2014 and from July 2018 to October 2019 have hindered continuous periodic analysis.
As a result, there have been no direct, continuous daily measurements of the rigidity dependence of the 9-day, 13.5-day, and 27-day periodicities during these periods, covering a broad range of rigidities.

Furthermore, the resolution of daily AMS measurements is insufficient for analyzing short-term cosmic ray variations, such as FDs.
Hence, measurements with higher temporal resolution (e.g., hourly data) are required to investigate these short-term cosmic ray variations.
More importantly, NMs provide continuous measurements at higher temporal resolutions.

In this study, we employ deep learning techniques to investigate the intrinsic relationship between NM data and AMS data, thereby enabling the calculation of proton flux from NM data over the period from 2011 to 2024.
This study is divided into two primary phases: NM data imputation and proton flux calculation.

In the NM data imputation phase, given that count rates from individual NM stations are independently and identically distributed, missing values often arise due to various factors, such as instrumental failures or upgrades to the NMs.
A comprehensive pre-processing pipeline consisting of multiple steps is initially applied to the raw NM count rates, including obtaining high-resolution corrected NM count rates, applying robust statistical outlier detection, computing daily averaged count rates from the dataset after filtering, and performing cross-checking of physically significant events across multiple NM stations.
Consequently, the missing values of pre-processed NM count rates are imputed using four advanced time series imputation algorithms.
Specifically, based on the experimental results and performance comparison introduced in PyPOTS \citep{du2023pypots}, SAITS \citep{du2023saits}, TimesNet \citep{wu2022timesnet} and Transformer-based methods demonstrate promising performance.
Therefore, we employ these deep learning models, which are trained and optimized on the pre-processed NM data, to further impute missing values.

In the proton flux calculation phase, leveraging the robust feature extraction capabilities of deep neural networks (DNNs), we propose and train a deep residual neural network to learn the features of the imputed NM data, aiming to predict the daily proton flux in the AMS dataset across various rigidity intervals.
Experimental results demonstrate that the trained deep model effectively captures the underlying patterns between NM count rates and AMS proton flux, achieving an $R^2$ score of 0.9984 on the test set.

In terms of application, since NMs provide high-resolution input to the proposed DNN, our approach can compute high temporal resolution cosmic proton flux, such as hourly flux, by utilizing the corresponding high temporal resolution NM dataset,
 which is important for studying rapidly changing transient events, such as FDs.

The unusual polar field reversal during Solar Cycle 24 is a critical phenomenon for understanding the dynamics of solar magnetic fields and their hemispheric asymmetries.
However, AMS measurements during this period suffer from data gaps, which hinder comprehensive analysis and prevent the application of wavelet analysis to study periodicities, as done for other continuous time intervals.
To address this challenge, our work reconstructs the missing data for AMS from this period, creating a continuous dataset.
Thanks to the high accuracy of our proposed predicting method, this reconstruction enables more detailed and accurate studies of the unusual polar field reversal, including wavelet analyses, contributing significantly to the understanding of solar and heliospheric processes during this period.

This paper is organized as follows: the machine learning model is presented in Section \ref{sec3}, which covers the sources and properties of the NM and AMS dataset, as well as the approach to proton flux modeling.
The computed daily proton flux and further analysis are presented in Section~\ref{sec4}, including the accuracy of daily proton flux calculated in Section~\ref{sec:daily}, uncertainty estimation for monthly proton flux in Section~\ref{sec:monthly}, wavelet analyses in Section~\ref{sec:wavelet}, and hourly proton flux in Section~\ref{sec:hourly}.
Finally, Section \ref{sec5} summarizes our findings and suggests potential avenues for future research.

\section{Modeling The Relationship between NM and AMS Measurements}
\label{sec3}
\subsection{Data Set}
\subsubsection{Neutron Monitor counts}
\label{sec21}
The NM dataset used in this study is obtained from the Neutron Monitor Database (NMDB) \footnote{\url{https://www.nmdb.eu/}}, which compiles measurements from over 50 stations worldwide, covering a wide range of longitudes and latitudes.
NMs measure cosmic ray flux, which varies across stations due to differences in geomagnetic cutoff rigidity. 
The rigidity values for these stations range from 0.01 GV (e.g., Terre Adelie) to 16.8 GV (e.g., Princess Sirindhorn Neutron Monitor).
This study uses a dataset spanning from January 1, 2011, to August 1, 2024.

Although NMDB provides daily records, gaps and outliers in the dataset require additional pre-processing.
Thus, we collect count rates from each station at higher temporal resolutions (e.g., 10-minute intervals, 30-minute intervals) over this period, because the time interval to record count rates for each NM depends on the location.
To match the daily resolution of AMS measurements and ensure uniform time intervals, NM count rates are processed to derive daily counts.

Each station’s daily count rate is calculated by averaging valid data points, defined as recorded measurements after outlier removal using Interquartile Range (IQR).
Outliers related to solar activity, such as FDs and Solar Energetic Particles (SEPs), are retained by comparing data from stations with similar geomagnetic cutoff rigidity.
Relevant events are preserved to ensure significant signals are not mistakenly excluded as anomalies.
Further details on the NM data pre-processing steps are provided in Appendix \ref{appendixA}.

The processed daily records from each station are concatenated column-wise to construct the NM dataset.
To ensure the reliability of the NM dataset, stations with more than 30 consecutive days of missing values in the NM count-rate series are excluded from our experiments.
After the filtering process, 18 stations are included in the following analysis: \texttt{AATB}(Alma-Ata B), \texttt{APTY}(Apatity), \texttt{FSMT}(Fort Smith), \texttt{INVK}(Inuvik), \texttt{JUNG}(Jungfraujoch IGY), \texttt{JUNG1}(Jungfraujoch NM64), \texttt{LMKS}(Lomnicky Stit), \texttt{MXCO}(Mexico), \texttt{NAIN}(Nain), \texttt{NEWK}(Newark), \texttt{OULU}(Oulu)\citep{oulu}, \texttt{PSNM}(Princess Sirindhorn Neutron Monitor), \texttt{PWNK}(Peawanuck), \texttt{SOPB}(South Pole 12-Bares), \texttt{SOPO}(South Pole), \texttt{TERA}(Terre Adelie), \texttt{THUL}(Thule) and \texttt{YKTK}(Yakutsk).

After applying the pre-processing and filtering procedures, short-term gaps still remain at 18 NM stations.
To address these gaps, we employ four advanced time series imputation algorithms to estimate the missing values.
Further details are provided in Appendix \ref{sec22}.

\subsubsection{Proton fluxes measured by the Alpha Magnetic Spectrometer }
\label{sec31}

The Alpha Magnetic Spectrometer (AMS) is a particle detector installed on the International Space Station (ISS).
Since its installation in May 2011, it has accumulated 13 years of cosmic ray records.
The dataset used in this study includes proton flux measurements collected by AMS from May 20, 2011, to October 29, 2019, spanning 8.5 years, corresponding to 2,824 days or 114 Bartels rotations (BR: 27 days) \citep{2021PhRvL.127A1102A}. AMS recorded a total of ${5.5 \times 10^9}$ protons, with flux measurements spanning from 1.00 to 100 GV across 30 rigidity ranges.
The dataset covers the ascending, maximum, and descending phases of solar cycle 24, offering a comprehensive view of cosmic ray behavior throughout the cycle.

The daily proton flux measurements used in this study are available on the AMS website \footnote{\url{https://ams02.space/publications/202105}}.
This AMS dataset is the first to conduct a periodicity analysis across multiple rigidities, including 9-day, 13.5-day, and 27-day periodicities.
To ensure the accuracy of this dataset, the AMS collaboration excluded measurements affected by SEPs with rigidity below 3 GV (from 1.00 to 2.97 GV) across 9 rigidity bins.
Additionally, some dates in the published flux measurements are missing due to detector studies and upgrades \citep{2021PhRvL.127A1102A}.

\begin{figure*}[htp]
    \centering
    \includegraphics[width=\linewidth]{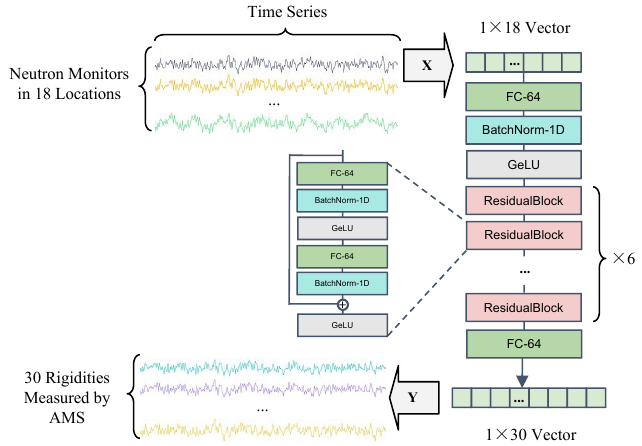}
    \caption{Illustration of the workflow for proton flux calculated and the architecture of the CNN.}
    \label{fig:nmams}
\end{figure*}

\subsection{The Proton Flux Model Based on Detected Neutrons}
\label{sec32}

The cosmic-ray proton flux is intrinsically correlated with NM counts: high-energy protons interacting in the atmosphere produce secondary showers, which in turn generate nucleons detected by NMs.
However, NM counts are also modulated by external factors such as atmospheric pressure, particle‐interaction cross‐sections, and shower characteristics.
In our experiments, we observe that deep time series models underperform due to gaps in AMS measurements while simplified Convolutional Neural Network (CNN) architectures with variable depths fail to converge due to vanishing gradients. 
By contrast, we found that residual blocks in CNNs demonstrate superior predictive performance and they consistently obtain stable convergence with the missing values problem.
To account for these complex interactions and to simplify the calculation of the relevant processes, we employ CNN with residual blocks.
Motivated by these findings, we adopt a residual‐block CNN to model the nonlinear mapping between cosmic-ray proton flux and NM counts.
This architecture efficiently extracts meaningful spatio‐temporal patterns while accounting for external covariates (e.g. air pressure and cross-sectional variations) that influence the measurements.

In this study, we aim to establish the relationship between NM count rates and proton flux on AMS.
Therefore, we align the imputed NM dataset with AMS proton flux by date, thereby creating a paired NM-AMS dataset that serves as inputs and outputs for training a calculated model.
More details on how to perform time series imputation on NM count rates and obtain the complete imputed NM dataset are presented in Appendix \ref{sec2}.
Specifically, the input features encompass complete NM count rates from May 20, 2011, to October 29, 2019, featuring 18 input variables, each corresponding to data from an individual NM station.
The model output represents the daily proton flux measured by AMS across various rigidity intervals for the same time period.

At first, for each day, we arrange the input features according to the rigidity of each NM station to form a $1\times18$ input vector as the input to a fully connected (FC) layer.
The FC layer linearly maps inputs from the original feature space to a 64-dimensional representation, thereby expanding the feature space and providing a consistent input dimension for subsequent deep residual blocks.
The reason for this dimensionality expansion is to maintain consistency with the Residual Block's input, which benefits from a higher-dimensional representation for better feature learning.
Subsequently, the batch normalization (BN) layer \citep{ioffe2015batch} is utilized to reduce the internal covariate shift by standardizing the intermediate activations within each training epoch, thereby enhancing the robustness and stability of the network.
In the deep feature extraction stage, we design six residual layers for feature extraction, each utilizing two FC layers, two 1-Dimensional BN layers, and two Gaussian Error Linear Units (GeLU) \citep{hendrycks2016gaussian} as the activation functions.
Those residual blocks \citet{he2016deep} are repeatedly employed to mitigate the vanishing gradient problem and improve the model’s learning capacity.
The use of residual connections enables the construction of deeper network architectures while maintaining training stability and efficiency.
In addition, an FC layer is incorporated for subsequent processing and outputs a  $1\times30$ vector.
The Residual Block's output vector is then projected back to 64 dimensions, allowing for a more effective dimensional reduction process before the final output is mapped to the desired 1x30 vector.
The architecture of the proposed deep residual neural network and the workflow for proton flux calculated are illustrated in Figure \ref{fig:nmams}.

In the experiment, we adopt an early stop strategy \citep{yao2007early} to further prevent overfitting.
The paired NM-AMS dataset is randomly partitioned into three subsets, with 80\% for the training set, 10\% for the validation set, and 10\% for the test set.
Records from a single day are treated as one sample for training, validation, and testing.
We set the batch size to 128 and the initial learning rate to $1\times10^{-3}$.
To optimize training, a learning rate decay strategy is employed.
In practice, if validation performance does not improve for 30 consecutive epochs, the learning rate is reduced by a factor of 0.2.
This reduction is repeated until the rate reaches a floor of $1\times10^{-9}$, after which no further adjustments are made.

We evaluate the performance of the time series imputation methods using Mean Absolute Error (MAE), Mean Percentage Error (MPE) and Root Mean Square Error (RMSE).
MAE, MPE, and RMSE are calculated using the formulas:
\begin{equation}
   \text{MAE} = \frac{1}{n} \sum_{i=1}^{n} |y_i - \hat{y}_i|,
   \label{eq:mae}
\end{equation}

\begin{equation}
   \text{MPE} = \frac{1}{n} \sum_{i=1}^{n} \frac{\hat{y}_i - y_i}{y_i} \times 100\% ,
   \label{eq:mpe}
\end{equation}

\begin{equation}
   \text{RMSE} = \sqrt{\frac{1}{n} \sum_{i=1}^{n} (y_i - \hat{y}_i)^2},
   \label{eq:rmse}
\end{equation}
where \( y_i \) represents the true value and \( \hat{y}_i \) denotes the calculated value for the \( i^{th} \) observation.
In addition to MAE, MPE and RMSE, we consider the coefficient of determination, $R^2$, to evaluate the accuracy of proton flux calculated, and is defined by the formula:
\begin{equation}
   R^2 = 1 - \frac{\sum_{i=1}^{n} (y_i - \hat{y}_i)^2}{\sum_{i=1}^{n} (y_i - \bar{y})^2}
   \label{eq:r2}
\end{equation}
where \( \bar{y} \) is the mean of the observations, indicating how well the calculated values approximate the actual measurements.
A higher \( R^2 \) value signifies a better fit of the model to the observations, with values ranging from 0 to 1, where values closer to 1 denote superior model performance.

Besides, we utilize chi-squared per degree of freedom ($\chi^2$/d.o.f.) to ascertain the accuracy of the trained model on the test set.
The chi-squared is defined as:
\begin{equation}
\chi^2 = \sum_{i=1}^{K} \frac{(\hat{y}_i - y_i)^2}{\sigma_i^2}
\end{equation}
where $\hat{y}_i$ is the predicted value, $y_i$ is the measured value, $\sigma_i$ is the uncertainty of bin $i$, and $K$ is the number of valid bins.
Define the reduced chi-squared as
\begin{equation}
\chi^2_{\text{d.o.f.}} \equiv \frac{\chi^2}{\text{d.o.f.}},
\end{equation}
where the number of degrees of freedom is $\text{d.o.f.} = K-p$ with $K$ the number of independent bins entering the sum and $p$ the number of model parameters estimated from the same data used to compute $\chi^2$.
In this study, no per-day parameters are fitted ($p=0$), meaning that the model's weights and hyperparameters remain fixed during the testing phase on the test set, and therefore $\text{d.o.f.} = K$.

\section{Results}
\label{sec4}

\subsection{The Accuracy of The Proton Flux Model}
\label{sec:daily}


\begin{figure}[h]
    \centering
    \includegraphics[width=0.5\linewidth]{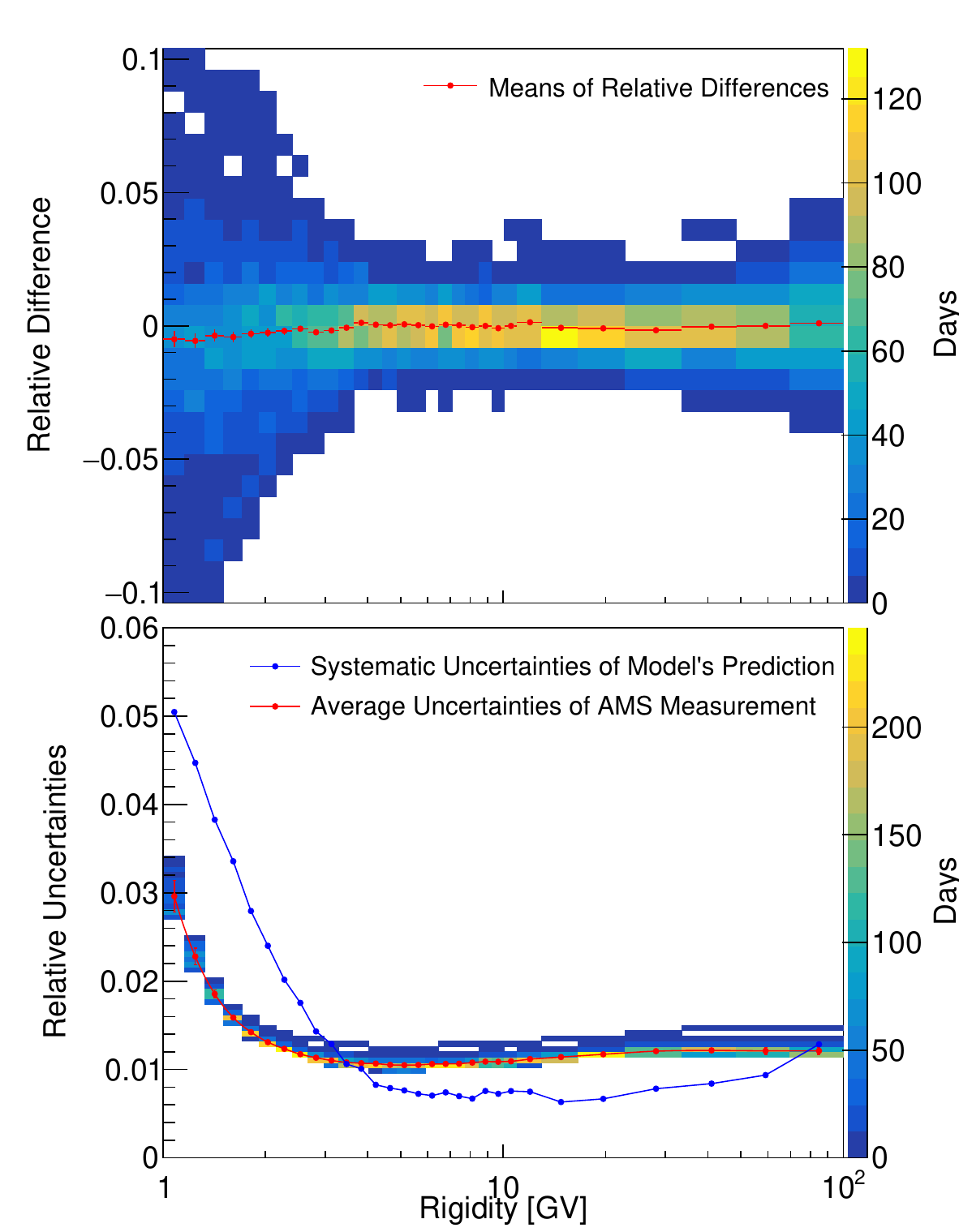}
    \caption{Comparisons between the model and the testing sample from the AMS measurement. The color shows the number of days. The top panel shows the relative difference between the model and AMS, where red points are the mean values in each rigidity. The bottom panel shows the relative uncertainties. The colored plot shows the relative uncertainty distribution of the AMS testing sample, while the red points are the mean values. The blue points are defined as the standard deviation of the relative difference in each rigidity bin from the top panel.
    }
    \label{fig:Predicted_to_Observed}
\end{figure}

In this experiment, our CNN model achieves an $R^2$ of 0.998 on the test set, indicating that the model is capable of accurately capturing the relationship between the ground-based NM dataset and the proton flux from AMS.
Additionally, the MAE is 4.950, the MPE is -0.104\%, the RMSE is 10.151 and the $\chi^2$/d.o.f. is 1.371 on the test set.
According to the MPE results, we observe that the model maintains good overall prediction consistency with very small deviation, demonstrating the model's accuracy and stability.
The near-zero MPE value further indicates that the model exhibits minimal systematic bias, suggesting that it neither consistently over- nor under-predicts the proton flux across the entire energy spectrum.
Meanwhile, the negative MPE indicates that the overall predicted values are slightly smaller than the measured values, which aligns with the results shown by the red line in the upper panel of Figure~\ref{fig:Predicted_to_Observed}.
The $\chi^2$/d.o.f. is close to 1, indicating a consistent prediction performance without significant overfitting.

In addition, the robustness of the residual block plays a significant role in ensuring that the performance results are relatively insensitive to the depth and dimensionality of the layers.
Specifically, the use of residual blocks helps mitigate the challenges of vanishing gradients and ensures effective feature propagation, making the model more stable and less dependent on the network depth or the number of neurons in the fully connected layers.
Our experiments have shown that different numbers of residual block layers and hidden units do not significantly affect the performance.
For example, when using 3 layers of residual blocks, the test set MPE is -0.109\%, while using 9 layers only result in a slight increase in MPE to -0.112\%.
Similarly, when varying the number of hidden units, we found that using 128 hidden units per layer led to an $R^2$ of 0.9982, and increasing this to 1024 hidden units resulted in a slightly lower $R^2$ of 0.9970.
On the other hand, reducing the number of hidden units to 32 still yields a respectable $R^2$ of 0.9974.
These similar results suggest that the model's performance is relatively stable across different network configurations, which further highlights the effectiveness of the architecture and the residual connections.

By applying our proposed model, we reconstruct a continuous AMS daily proton flux dataset covering 2011–2024.
This dataset fills gaps arising from AMS observational interruptions during detector upgrades and extends the record beyond the officially published daily AMS proton flux to include the post-2019 period.
To assess the predictive accuracy of our method, we compute the relative deviation of the model-calculated proton flux from the AMS measurements within multiple rigidity bins, defined as:
\begin{equation}
   \text{Relative Difference}=\left(\frac{\text{Flux}_{\text{model}}}{\text{Flux}_{\text{AMS}}} - 1\right).
   \label{eq:pfpp}
\end{equation}
As is shown in Figure~\ref{fig:Predicted_to_Observed}, the upper panel shows the relative difference between the model and AMS measurement as a function of Rigidity calculated from the testing sample.  The colored histogram shows the number of days for a specific relative difference in a rigidity bin, with red points indicating the mean values of the relative difference.
The lower panel shows the relative uncertainties as a function of Rigidity. The colored histogram, showing the number of days, is the distribution of the relative uncertainties calculated from the quadratic sums of AMS  ``statistical'' and ``time-dependent'' error. The red points indicate the mean values of AMS relative uncertainties. The blue points connected by a line represent the standard deviations of the relative differences shown in the upper panel, which we assign as systematic uncertainties of our model.
Overall, the results demonstrate good agreement between the model predictions and the AMS measurement across a wide rigidity range, with small discrepancies in some bins that likely arise from flux variations and AMS measurement uncertainties.

Moreover, the distribution of daily calculation errors across rigidity bins approximates a Gaussian. We compute calculation errors on the AMS daily test subset spanning May 2011–November 2019 to assess model accuracy. The lower panel of Figure~\ref{fig:Predicted_to_Observed} shows relative systematic errors: blue dots are individual daily errors, the overlaid histogram represents AMS time-dependent uncertainties, and red markers denote mean error per rigidity bin. The close correspondence of these patterns across all rigidity bins confirms the robustness of our approach.

\begin{figure}[h]
    \centering
    \includegraphics[width=0.75\linewidth]{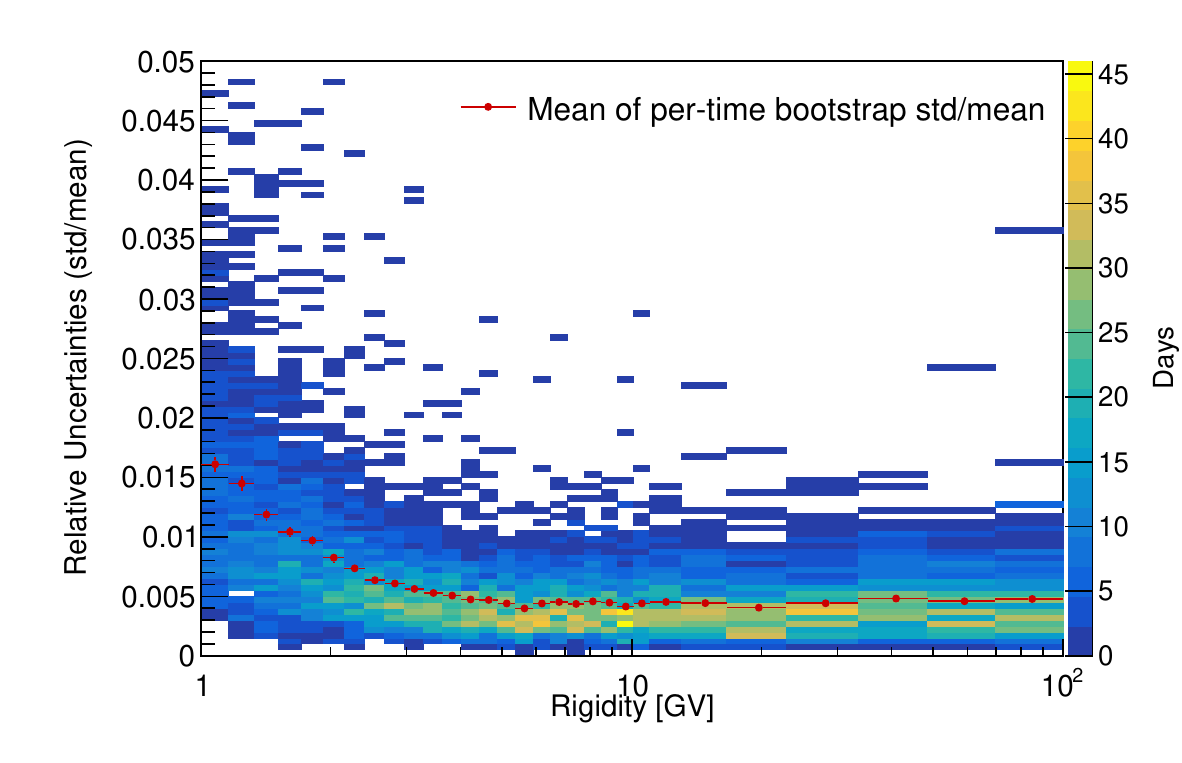}
    \caption{Bootstrap-based relative uncertainties of the reconstructed daily proton flux versus rigidity. This plot describes the variations of the model.
    The color plot shows the number of days at a given rigidity and relative uncertainty bin.
    The red points are the mean relative uncertainties.
    }
    \label{fig:time_std_vs_ams}
\end{figure}

In addition, we assess the model’s uncertainty via bootstrapping over training/validation splits.
We train five independent replicas with different random seeds while keeping the same test set.
For each day and rigidity bin, we compute the relative dispersion across replicas. 
Figure~\ref{fig:time_std_vs_ams} shows the number of days for a given std/mean and rigidity.
The mean values shown in red show the rigidity dependence of the relative uncertainty.
Compared to the systematic uncertainties in the lower panel of Figure \ref{fig:Predicted_to_Observed}, the bootstrap curve in Figure \ref{fig:time_std_vs_ams} shows the same rigidity trend.

\subsection{Analysis of Monthly Proton Flux}
\label{sec:monthly}

The AMS collaboration has released monthly proton flux, aggregated over intervals defined by BRs and with wider rigidity bins compared to the daily dataset \citep{2021PhRvL.127A1102A}.
These monthly datasets provide extended observational coverage, particularly from November 2019 to June 2022, extending beyond the time span of the officially released daily records \citep{PhysRevLett.134.051002}.
To evaluate the accuracy and robustness of the model at different temporal resolutions, we apply our model to the monthly measurements.
Specifically, we first aggregate ground-based NM count rates using the start and end times of each BR to produce time-aligned inputs.
These are then fed into our model to predict the space-based proton flux at each BR interval.

\begin{figure*}[ht]
    \centering
    \includegraphics[width=1\textwidth]{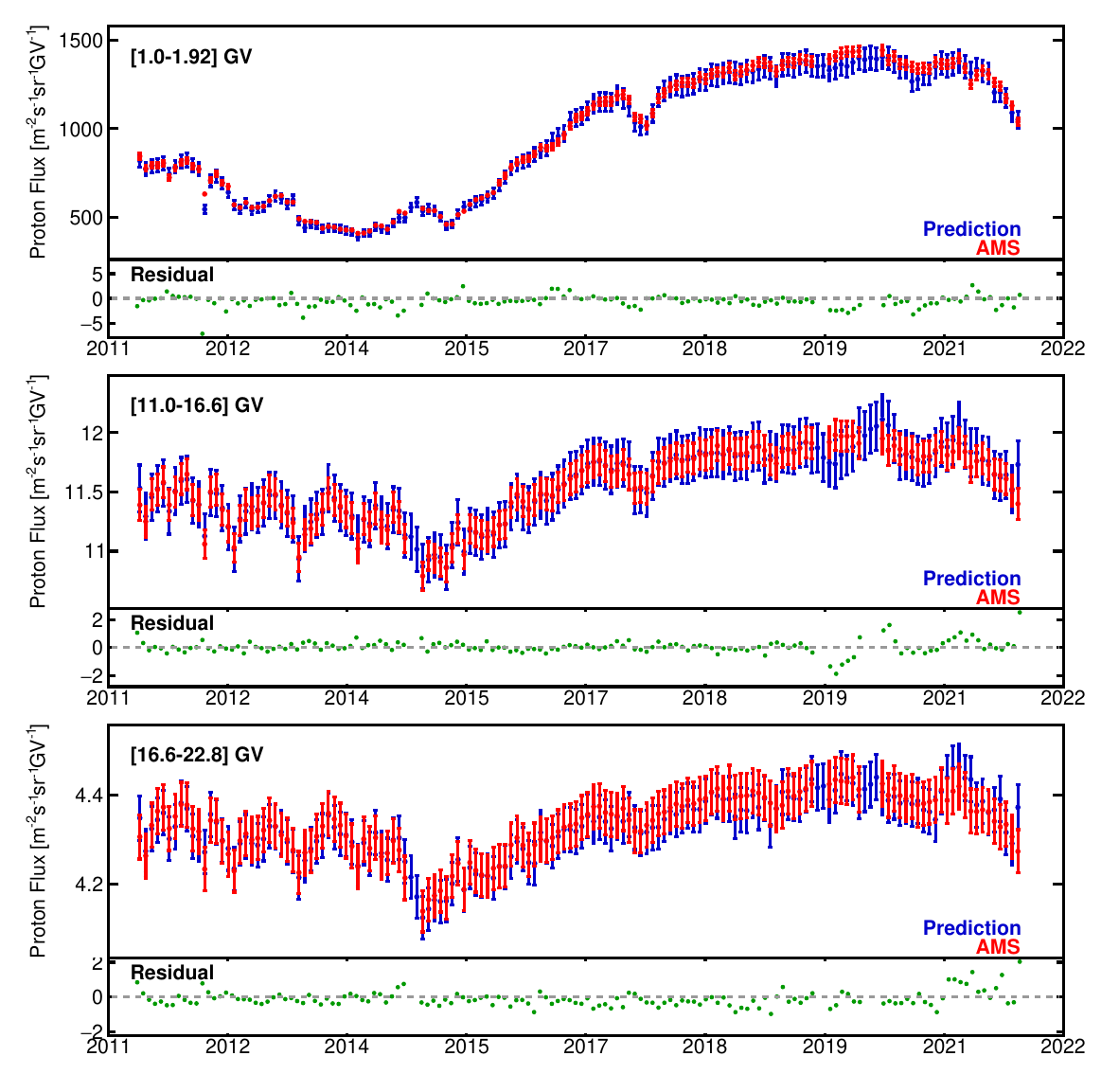}
    \caption{Model calculations aggregated at BR resolution are compared with AMS proton flux measurements. Three representative rigidity intervals, [1–1.92] GV, [11–16.6] GV, and [16.6–22.8] GV, are consistent with the AMS measurement. We define the residual as $\mathrm{Residual}=({\Phi_{\mathrm{NM\enspace Prediction}}-\Phi_{\mathrm{AMS}}})/{\sigma_\mathrm{{AMS}}}$, residuals remain small before 2019 and increase afterward. The error bars on the monthly prediction results represent a conservative estimate of the systematic deviation.}
    \label{fig:compared_with_model_br}
\end{figure*}

The output of the model, initially generated in 30 finite rigidity bins, is subsequently combined into a wider rigidity interval consistent with the monthly AMS dataset.
This matching ensures a one-to-one comparison with the published AMS values on the time and rigidity dimensions.
The purpose of this comparison is to validate the robustness of the model over extended time periods and to take advantage of the richer temporal coverage provided by the AMS monthly measurements in recent years. Figure~\ref{fig:compared_with_model_br} shows the comparison results in selected rigidity intervals, which shows a good general agreement between the model-computed flux and the AMS measurements.

In each selected typical rigidity bin of Figure~\ref{fig:compared_with_model_br}, the model-calculated flux is consistent with the AMS measurements.
However, at lower rigidities, the level of agreement is reduced compared to that at higher rigidities.
In the post-2019 period, the AMS measurements have not been used in the machine learning process. The discrepancies between our model and AMS become larger than those in the pre-2019 period.
These discrepancies are treated as additional systematic uncertainties associated with the model.
To quantify these model uncertainties, we define an effective relative error $\epsilon$ for each rigidity bin.
The calculation is performed using measurements from November 2019 to ensure sufficient monthly coverage.
At each point, the absolute difference between the calculated and measured flux is computed, and the AMS total uncertainty is subtracted.
Only deviations exceeding the AMS uncertainty are considered:
\begin{equation}
\epsilon = \max \frac{\left| F_{\text{pred}}(t) - F_{\text{AMS}}(t) \right| - \sigma_{\text{AMS}}(t)}{F_{\text{pred}}(t)}.
\label{eq:epsilon}
\end{equation}

where $F_{\text{pred}}(t)$ is the flux predicted by the model, $F_{\text{AMS}}(t)$ is that measured by AMS, and $\sigma_{\text{AMS}}(t)$ is the uncertainty of AMS measurement. The maximum value of $\epsilon$ over all selected time points is adopted as the representative relative error for each rigidity bin. 
The plotted uncertainties are then obtained by multiplying the calculated flux by this maximum relative error.
This method provides a conservative estimation of the model deviation, considering only discrepancies that exceed the experimental uncertainties. 
We select this method to estimate the error because it provides a more reliable upper limit on the model's calculation error.

\begin{figure}[ht]
    \centering
    \includegraphics[width=\linewidth]{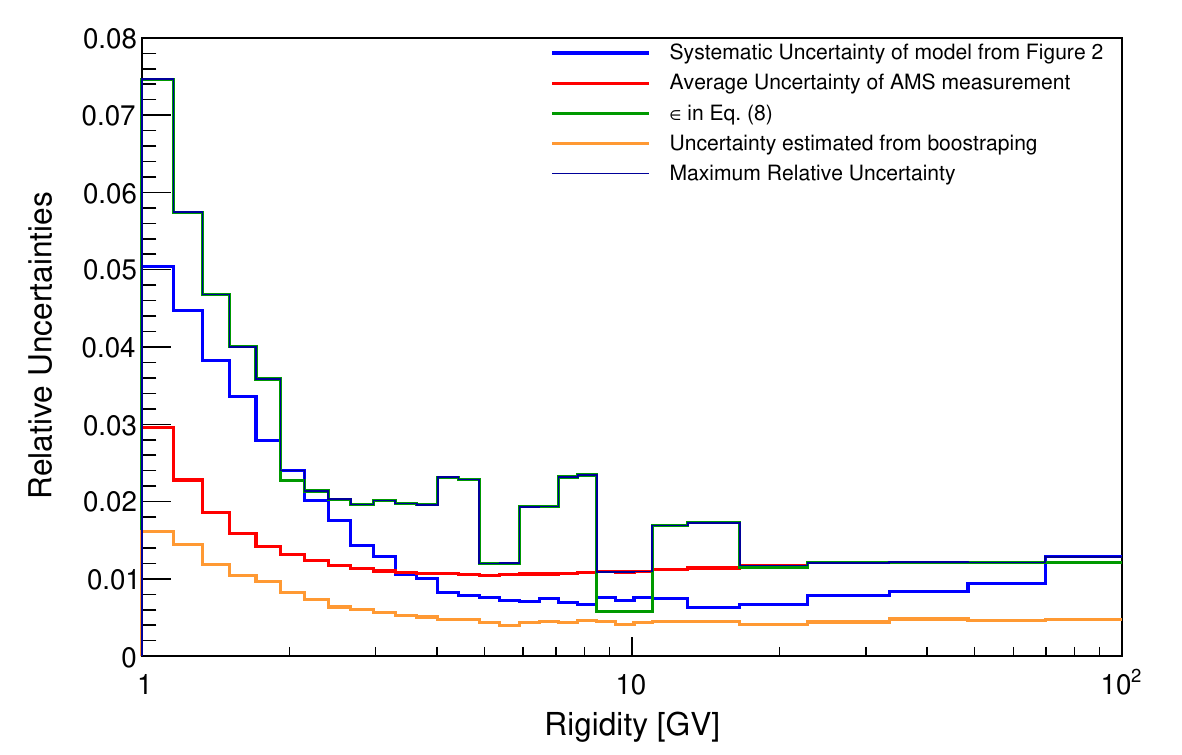}
    \caption{Total error estimation for the model. The black dashed line shows the overall model uncertainty, conservatively defined as the maximum of three distinct error sources: systematic uncertainty of model from Figure~\ref{fig:Predicted_to_Observed}(blue), average uncertainty of AMS measurement(red) and $\epsilon$ in Eq.~\ref{eq:epsilon}(green). The green line illustrates the post-2019 model error component, estimated based on the rigidity dependence of daily AMS measurements. For comparison, the orange line shows an independent uncertainty obtained by bootstrapping.}
    \label{fig:errors}
\end{figure}

Finally, the total error is defined as the maximum value from three sources: AMS measurement errors, model errors (pre-2019), and model errors (post-2019).
Since the rigidity bins of the AMS daily and BR measurements do not align, the post-2019 model errors are estimated by interpolating the BR-derived errors to match the rigidity bins of the daily AMS measurements.          
The estimation is shown as the green line in Figure~\ref{fig:errors}, while the black dashed line represents the maximum error from the three sources.
This approach provides a conservative estimate of the uncertainties by defining an upper bound based on the largest value among the three error sources.

\subsection{Wavelet Analysis of Daily Proton Flux}
\label{sec:wavelet}

\begin{figure*}[ht]
    \centering
    \includegraphics[width=1\textwidth]{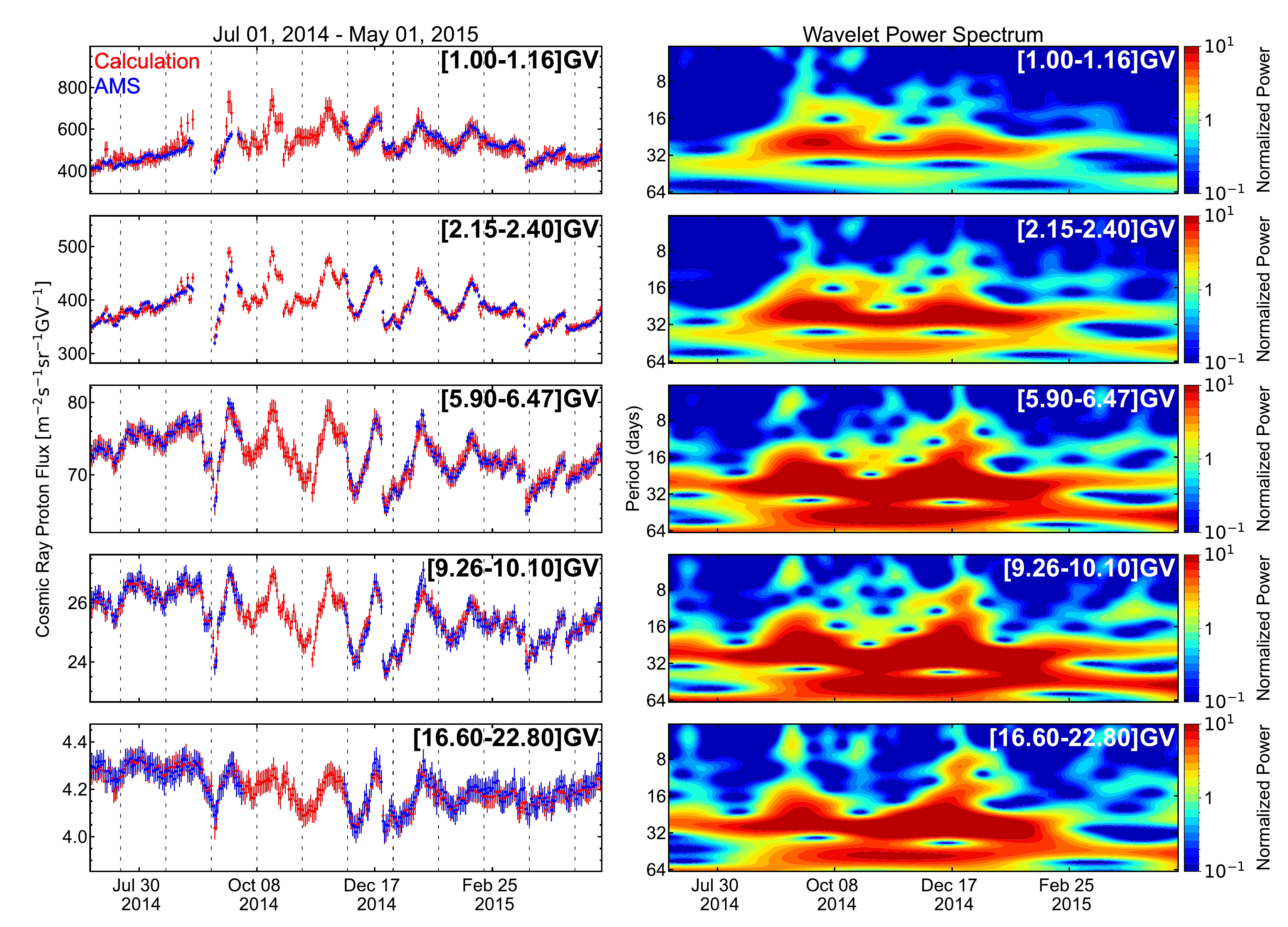}
    \caption{The left plot shows the daily proton flux for five different rigidities, ranging from low to high, between 1 July 2014 and 1 May 2015. The proton flux, which we calculate from the ground NM station (in red), fills the gaps in the AMS measurement (in blue), particularly during the reversal of the solar magnetic field, Vertical dashed lines indicate the boundaries of the BR. In order to maintain consistency with the reporting results of the AMS (see \citet{2021PhRvL.127A1102A}), when a SEP event occurs, the proton flux below 3 GV has been excluded from the relevant low rigidity interval. The right plot shows the wavelet time-frequency power spectrum corresponding to the same rigidity ranges as the left plot, the color scale represents the normalized power.}
    \label{fig:last2wavelet}
\end{figure*}

Cosmic ray flux is modulated by various physical processes such as solar activity, interplanetary magnetic fields, and heliospheric shielding effects. It usually exhibits complex time-varying characteristics and contains a variety of periodic components. The amplitude, phase, and frequency of these cycles are often not constant, but vary with time, especially at different stages of the solar activity cycle.

This non-stationary feature makes traditional Fourier analysis methods significantly limited.
Although the Fourier transform can decompose the signal into a superposition of frequency components, it cannot provide temporal information on when these frequency components appeared and how they evolved. 

Wavelet analysis is an effective tool to solve this problem. 
In this study, we apply the wavelet analysis techniques \citep{APracticalGuidetoWaveletAnalysis}, which provides a comprehensive framework for analyzing non-stationary time series across multiple timescales.
In our analysis, the time series \( X_t \) represents the proton flux \( x_n \) at each time index \( n \), with the measurements sampled in a constant time interval \( \delta t \), corresponding to one day. The wavelet transform \( W_n(s) \) is then computed as the convolution of the wavelet function $\psi$ with the proton flux time series \( x_n \):

\begin{equation}
    W_n(s) = \sum_{n'=0}^{N-1} x_{n'} \psi^* \left[ \frac{(n' - n) \delta t}{s} \right]
    \label{eq:wavelet_transform}
\end{equation}
where wavelet function $\psi$ is defined as 

\begin{equation}
\psi \left[ \frac{(n' - n) \delta t}{s} \right] = \left( \frac{\delta t}{s} \right)^{1/2} \psi_0 \left[ \frac{(n' - n) \delta t}{s} \right]
\end{equation}
and the asterisk (*) indicates the complex conjugate.
$\psi$ is a scaled and time-shifted form of the mother wavelet $\psi_0$, which will be defined later. The scale is adjusted by the dilation parameter $s$, and the wavelet is shifted in time according to the translation parameter $n$. The factor $s^{1/2}$ is used as a normalization to keep the total energy of the scaled wavelet constant, ensuring that the shape of the wavelet remains consistent while its size changes with the scale.

In this work, the mother wavelet $\psi_0$ we use is the Morlet wavelet, which is defined as the product of a complex exponential wave and a Gaussian envelope:
\begin{equation}
    \psi_0(\eta) = \pi^{-1/4} e^{i\omega_0 \eta} e^{-\eta^2/2}
    \label{eq:wave_function}
\end{equation}
where \( \eta \) is the non-dimensional time, and \( \omega_0 \) is the wave number, and we set \( \omega_0 \) to 6 in our experiments. 

We perform wavelet analysis on proton flux in five representative rigidity bins: 1.00-1.16 GV, 2.15-2.40 GV, 5.90-6.47 GV, 9.26-10.10 GV and 16.60-22.80 GV.
Before analysis, each time series is standardized with Z-score normalization by subtracting the mean of the series and dividing by its standard deviation.
Z-score normalization standardizes the proton flux across different rigidity bins, removing the effect of varying magnitude scales.
This standardization ensures consistent amplitude scaling in the wavelet analysis, thereby enabling direct comparison of power spectra between rigidity ranges.

Initially, we calculate the periodicity for the entire dataset, ranging from 2011 to 2024.
Since we are particularly interested in the periodicity during the exceptional polar field reversal of Solar Cycle 24, we primarily present the analysis results for this specific period, as shown in Figure \ref{fig:last2wavelet}.
The left panel of Figure~\ref {fig:last2wavelet} shows our calculated flux can match AMS measurements for all rigidity ranges, which is significantly better than the results calculated from the traditional NM force-field reconstruction methods by \cite{2025JGRA..13033805V}, where they obtained good results only between 5-15 GV. 

The results of the wavelet analysis are visualized by plotting the time series of proton flux alongside the wavelet power spectrum.
For each rigidity bin, we plot the proton flux time series along with the global wavelet power. This approach enables us to visually compare the oscillatory behavior across different energy levels and periods, providing valuable insights into the time-varying characteristics of cosmic ray flux.

Our analysis, especially for the period of solar magnetic pole reversal in the 24th solar cycle, shows that the observed periodic behavior of proton flux is basically consistent with that in other periods.
We did not find that this magnetic pole reversal event showed other periodicities.
The analysis results still show that at lower rigidities, the periodicity related to solar rotation (about 27 days) dominates; while as rigidity increases, the shorter periodicities of about 13.5 days and 9 days gradually become more significant.
These findings are generally consistent with the observations previously reported by the AMS Collaboration\citep{2021PhRvL.127A1102A}, suggesting that even during the reversal of the solar magnetic field, there may be a stable physical mechanism behind these periodicities.
\subsection{Hourly Proton Flux \& Forbush Decrease}
\label{sec:hourly}

In addition, we also apply the model to hourly proton flux.
Since there is a diurnal cycle variation in the hourly NM count rates on the ground \citep{tiwari2012study}, which is not reflected in the space data.

we apply the Seasonal-Trend decomposition using Loess (STL) \citep{cleveland1990stl} with a 24-hour seasonal period. By removing the estimated diurnal component, we obtain a diurnal–cycle–free series. The time-varying amplitude of the 24-hour cycle has already been accounted for by the STL method.
In this way, the hourly proton flux is further used for cosmic ray studies, such as the study of the FDs in the next paragraph.
However, to the best of our knowledge, we did not find any publication about hourly proton flux above 1.0 GV, so we are currently unable to verify the accuracy on the hourly resolution.

FDs are sudden decreases in GCRs due to intense solar wind transients, which typically occur over several hours or days.
They are primarily caused by Interplanetary Coronal Mass Ejections (ICMEs) and Corotating Interaction Regions (CIRs). In the case of ICMEs \citep{2017SSRv..212.1159M}, when a bulk of plasma is ejected from the Sun, it can lead to variations in cosmic ray densities. This type of cosmic ray decrease typically features a sudden drop and a more gradual recovery, resulting in an asymmetric profile with time. Moreover, if the ejecta is energetic enough to generate a shock and the observer is in the path of the ejecta, a two-step FD will be observed. There are three types of FDs: shock and ejecta, shock only, and ejecta only \citep{2000SSRv...93...55C}. Most decreases fall into the first category. 
The period for these two-step FDs, where the first and second steps correspond to the arrivals of the shock and the ejecta-shock boundary, respectively, can be within one day. Consequently, daily flux is insufficient to resolve the structure of such FD events. However, hourly flux measurements offer an opportunity for identifying and analyzing these two-step features.

The other type of FDs, caused by solar wind interactions known as CIRs \citep{2018LRSP...15....1R}, occurs when solar wind originating from a coronal hole catches up to and compresses the slower wind that was emitted earlier. In this compressed plasma region, increased turbulence can prevent cosmic rays from entering. Additionally, the fast wind can sweep away cosmic rays from the solar wind region. The time profile of CIR-related FDs is symmetrical. Unlike daily flux measurements, hourly resolution measurements can effectively distinguish between FDs induced by ICMEs and those driven by CIRs. Since coronal holes rotate with the Sun, this region can persist for many months, sharing the Sun's 27-day rotational period. Such FDs can be utilized to analyze the evolution of coronal holes.

The research of FDs has long relied on ground-based NMs, which provide long-term observations. However, NMs measure the total flux after energy integration and cannot distinguish between particle species \citep{WAWRZYNCZAK2010622}.

In contrast, space detectors such as PAMELA, DAMPE, and AMS can directly detect different particle species and measure their energy spectra, which is more useful for studying FDs. However, the temporal resolution of currently published flux is low. For example, \cite{2023ApJ...950...23W} analyzed proton flux from AMS; in that study they investigated the evolution of the cosmic ray rigidity spectrum during FDs and explored the correlation between FDs amplitudes and solar wind parameters. However, the daily resolution measurements remains insufficient to analyze the more detailed structure of FDs.

\cite{2021ApJ...920L..43A} offers 6-hour resolution cosmic-ray electrons and positrons, enabling valuable analyses of FD amplitudes and recovery times, but the DAMPE experiment does not distinguish between positrons and electrons.
Thus, high temporal resolution measurements are is essential for addressing the current research gap and providing a more detailed analysis of FDs.

Our work provides such dataset with hourly resolution, enabling the community to capture the rapid changes during FDs more accurately, thereby facilitates a deeper understanding of the energy dependence. 
Our research provides a more detailed framework for comprehending the interactions between solar activities and cosmic rays. As a result, our provision of hourly proton flux makes a substantial contribution to the field by offering more precise tools for studying FDs.

\begin{figure*}[ht]
    \centering
    \includegraphics[width=\linewidth]{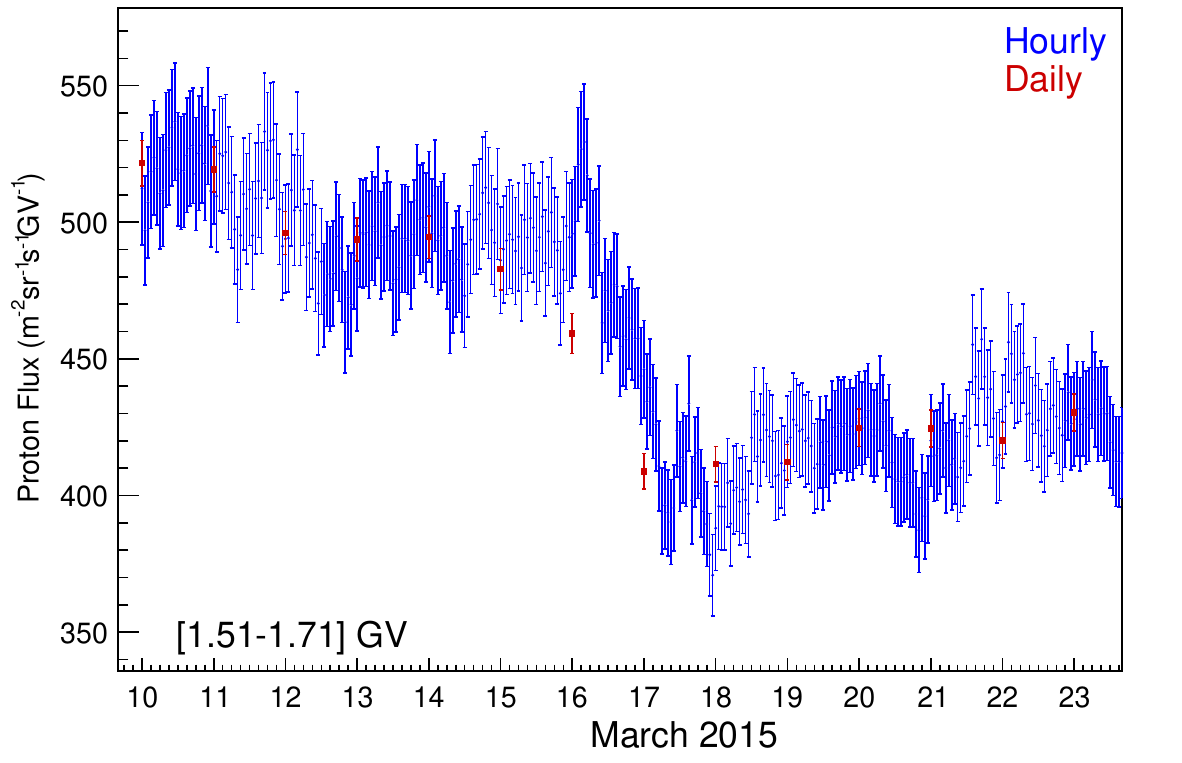}
    \caption{Absolute proton flux during the March 2015 Forbush Decreases, comparing AMS daily measurements (red points) with our model's hourly computed flux (blue points) for the rigidity interval 1.51–1.71 GV. Two FDs occurring on 16 and 17 March 2015 are both attributed to ICMEs. The higher temporal resolution of the calculated flux captures the rapid changes during this period more effectively than daily measurements, allowing detailed analysis of cosmic ray modulation on short timescales.}
    \label{fig:FDs}
\end{figure*}

Figure~\ref{fig:FDs} presents a comparison between the daily proton flux measurements reported by AMS and the hourly proton flux computed by our model during a solar activity. The yellow points represent the relative variations of the model-calculated proton flux at hourly resolution, while the red points show the corresponding relative variations in the daily AMS measurements. 
This solar activity resulted in two distinct FDs, both caused by ICMEs, occurring on 16 and 17 March 2015. The model-calculated higher temporal resolution enables more detailed observation of the rapid flux decreases and recoveries that are characteristic of FDs. This provides valuable insight into how solar transients dynamically modulate cosmic rays.

In summary, our research enhances the understanding of FDs by using hourly proton flux dataset. This approach enables a more detailed analysis of the dynamic changes during FDs, and we anticipate that these refined data will provide a solid foundation for further investigations into the complex interactions between solar activities and the cosmic-ray modulation.

\section{Conclusion}
\label{sec5}
In this study, we introduce a new method for measuring proton flux using count rates from ground-based NM stations combined with machine learning techniques. After preprocessing the NM count rates, we utilize a deep residual neural network to establish a correlation between NM observations and proton flux measurements from the Alpha Magnetic Spectrometer (AMS). With this model, we generate daily proton flux dataset covering the period from 2011 to 2024.

We assess the model's performance by comparing its results with AMS observations across 30 different rigidities in the test set. The close alignment between our model's measurements and the AMS proton flux measurements indicates that the convolutional neural network (CNN)-based approach effectively captures the relationship between NM signals and proton flux.

Additionally, we conduct wavelet analyses of the continuous proton flux dataset to explore the effects of solar activity. Specifically, we examine the unusual solar polar field reversal that occurred in 2014 (during Solar Cycle 24) and its potential impact on cosmic ray variations. At lower rigidity levels, the proton flux predominantly exhibits a dominant 27-day periodicity, which corresponds to solar rotation. In higher rigidity ranges (e.g., 16.6–22.8 GV), we observe additional shorter periodicities of approximately 13.5 days and 9 days. These findings align with previous results from AMS \citep{2021PhRvL.127A1102A}, suggesting that the magnetic field reversal did not significantly affect the fundamental periodic behavior of the proton flux. 

In addition, we validate the model's generalization capability by comparing its outputs with the updated AMS dataset, which extends to June 2022.
The strong agreement across various energy bands further confirms the robustness of our method.
Finally, acknowledging the importance of high temporal resolution for capturing short-term cosmic ray variability, such as FDs, we provide additional proton flux dataset at one-hour intervals.
This offers a valuable resource for detailed studies of rapid fluctuations in cosmic-ray flux.
The dataset can be accessed at Zenodo\footnote{\url{https://zenodo.org/records/17181237}}.

We acknowledge that there is always room for improvement, and we will consider exploring other network architectures in future work to further enhance model performance.
There are many encouraging time series models have been proposed in the field of time series forecasting recently, like Tsmixer \citep{ekambaram2023tsmixer} and DLinear \citep{zeng2023transformers}.
These time series models employ entirely different model designs and parameters compared to CNN architectures.
Unlike the CNN model used in this paper, time series models consider the correlation between continuous measurements along the time dimension to predict the proton flux in AMS.
The CNN model presented in this paper will serve as a benchmark for comparison in the subsequent work, providing valuable insights into the trade-offs between model complexity and performance.

\begin{acknowledgments}
This work is supported in part by the Guangdong Provincial Key Laboratory of Advanced Particle Detection Technology (2024B1212010005), the Guangdong Provincial Key Laboratory of Gamma-Gamma Collider and Its Comprehensive Applications (2024KSYS001), the Fundamental Research Funds for the Central Universities, and the Sun Yat-sen University Science Foundation.
We acknowledge the NMDB database \url{https://www.nmdb.eu}, founded under the European Union's FP7 programme (contract no. 213007) for providing data. 
We also thank the principal investigators and institutions of the following stations for maintaining and providing the data:
The neutron monitor data from AATB are provided by Institute of Ionosphere, Almaty, Kazakhstan (PI: Dr. O.N. Kryakunova);
The neutron monitor data from APTY are provided by Polar Geophysical Institute Russian Academy of Sciences, (PI: Boris Gvozdevsky);
The neutron monitor data from Fort Smith are provided by the University of Delaware Department of Physics and Astronomy and the Bartol Research Institute (PI: John Bieber);
The neutron monitor data from Inuvik are provided by the University of Delaware Department of Physics and Astronomy and the Bartol Research Institute;
Jungfraujoch neutron monitor data are kindly provided by the Physikalisches Institut, University of Bern, Switzerland;
The operation of the LMKS neutron monitors is supported by Slovak grant agency APVV, project no. 51-053805;
Mexico City neutron monitor data are kindly provided by the Cosmic Ray Group, Geophysical Institute, National Autonomous University of Mexico (UNAM), Mexico;
The neutron monitor data from Nain are provided by the University of Delaware Department of Physics and Astronomy and the Bartol Research Institute (PI: John Bieber);
The neutron monitor data from Newark/Swarthmore are provided by the University of Delaware Department of Physics and Astronomy and the Bartol Research Institute (PI: John Bieber);
The neutron monitor data from OULU are provided by Sodankyla Geophysical Observatory, University of Oulu, SGO-OTY, 90014 University of Oulu, Finland (PI: Ilya Usoskin);
Neutron monitor data from Doi Inthanon are provided by courtesy of the Princess Sirindhorn Neutron Monitor Program (PI: Prof. David Ruffolo);
The neutron monitor data from Peawanuck are provided by the University of Delaware Department of Physics and Astronomy and the Bartol Research Institute (PI: John Bieber);
The neutron monitor data from the South Pole Bares are provided by the University of Wisconsin, River Falls (PI: Paul Evenson);
The neutron monitor data from South Pole are provided by the University of Wisconsin, River Falls (PI: Paul Evenson);
Terre Adelie neutron monitor data are kindly provided by Observatoire de Paris and the French Polar Institute (IPEV), France;
The neutron monitor data from Thule are provided by the University of Delaware Department of Physics and Astronomy and the Bartol Research Institute (PI: John Bieber);
The neutron monitor data from YKTK are provided by the Institute of Cosmophysical Research and Aeronomy of Russian Academy of Science.
Python wavelet software provided by Evgeniya Predybaylo based on \cite{APracticalGuidetoWaveletAnalysis} and is available at \url{http://atoc.colorado.edu/research/wavelets/} .
We are grateful for the fruitful discussions about physics and code with Junhua Li and Yi Jia.
\end{acknowledgments}

\begin{contribution}




Jie Feng came up with the initial research ideas for this work led by his funded project and improved writing of the manuscript.
Pengwei Zhao was responsible for wavelet analysis, data collection and writing up the manuscript.
Jianqi Yan was responsible for the ideas and experiments for the machine learning methods, model training and testing, and writing up the manuscript.
Alex P. Leung provided insights in modeling and validation. He also improved writing of the manuscript.


\end{contribution}

%



\appendix

\section{Neutron Monitor Data Imputation}
\label{sec2}

\subsection{Neutron Monitor Data Imputation Methodology}
\label{sec22}

After the pre-processing described in Appendix \ref{sec21}, the preprocessed NM count rates still retains some missing values due to the downtime of the NMs.
These partially-observed time series can be a significant barrier to further analysis and modeling.
To address this issue, we employ four advanced time series imputation algorithms, SAITS \citep{du2023saits}, iTransformer \citep{liu2023itransformer}, TimesNet \citep{wu2022timesnet}, and Transformer \citep{vaswani2017attention}, to effectively reconstruct the missing values.
Notice that there is not any single day with the complete absence of records across all 18 stations, as they are supposed to provide redundancy for observations throughout the year.
We further exploit inter-station spatial-temporal patterns for reliable imputation.
With the efficient time series imputation methods provided by PyPOTS \citep{du2023pypots}, we compare the imputation performance of various models within a unified framework, subsequently selecting these four methods for in-depth performance and analysis.

To better understand their imputation performance, we next delve into the key ideas of four advanced time series imputation algorithms.
These four different network architectures selected have been used as benchmarks in multiple time series tasks due to their outstanding performance \citep{liu2024timer,wang2024deep}, hence these approaches are employed in this work.
The key idea of SAITS is to utilize the self-attention mechanism to accurately capture the complex interdependencies between different time steps in multivariate time series.
Even in environments with missing values, SAITS can realistically reconstruct the original data distribution.
In this context, the model simultaneously optimizes the reconstruction of observed data and the calculation of intentionally masked values.
This dual optimization ensures precise fitting of the visible data and embeds the capability to infer potential missing values within its deep feature representations.

Transformer \citep{vaswani2017attention} is a well-known sequence architecture in recent years.
It applies self-attention and position embeddings to model global dependencies across all time steps, offering feature extraction without explicit assumptions on periodicity.
As a widely adopted architecture, we also include it as one of the models for comparison in this work due to its outstanding design.
iTransformer, based on the Transformer architecture, aims to enhance feature extraction capabilities by learning dependencies among sequences.
However, iTransformer initially lacked the capability to directly handle partially-observed time series.
To overcome this limitation, the PyPOTS toolbox applies to iTransformer the same embedding strategy and training approach as SAITS, thereby enabling it to accept multivariate time series with missing values as input \citep{du2023pypots}.
In our experiments, we thus employ this PyPOTS-modified iTransformer variant, which retaining strong feature-extraction capabilities while supporting missing values imputation.

In addition to iTransformer, we also utilize the PyPOTS-modified version of TimesNet for comparison.
TimesNet is designed specifically for temporal data, excels in capturing long-term dependencies in time series through a hybrid deep learning architecture.
It combines convolutional layers with self-attention mechanisms to improve the model’s ability to learn both local and global patterns.
Furthermore, it captures multi-periodicity in the data by converting the 1D time series into a 2D format.
This allows the use of 2D convolutions to extract both intra-period and inter-period features.
By incorporating multi-scale convolutions, TimesNet can better capture and reconstruct temporal dynamics compared to traditional sequence models.
However, its dependence on predefined periodic bases limits its flexibility when dealing with irregular missing patterns or the nonstationary behaviors commonly observed in NM data.


With the pros and cons, we found that SAITS performs better in the experiments.
Further details of the experimental results are presented in  Appendix \ref{sec23}.

Before training, the original multivariate time series is divided into three datasets: 80\% for the training set, 10\% for the validation set, and 10\% for the test set.
It is noteworthy that, to further evaluate the models' generalization performance and robustness, we employ the Missing Completely At Random (MCAR) strategy to randomly mask 30\% of the observed values in the training, validation and test sets.
This procedure is also utilized in PyPOTS.
Specifically, for each training iteration, the random masked values are predicted by the model and the mean squared error between the ground truth and predicted value is further calculated on the validation set.
We then employ the trained imputation model to recover the masked entries on the test set, achieving accurate and reliable reconstructions.
Subsequently, the imputed NM dataset is then used as input to predict the daily proton flux in the AMS dataset across various rigidity intervals in the next phase.

As is shown in Table \ref{tab:Imputation1}, the percentage of the imputed NM dataset is less than 0.7\%, indicating that the imputed NM count rates does not significantly affect the statistical properties of the original time series.

\begin{table*}
\centering
\setlength{\tabcolsep}{6pt}
\renewcommand{\arraystretch}{1.2}
\caption{Numbers and percentages of daily records removed as outliers and filled by imputation are shown for each NM station. Outliers are detected with IQR (Section \ref{sec21}), and the imputed values correspond to dates that are originally missing or later removed as outliers. 577 out of 89,316 NM count rates are imputed.}
\begin{tabular}{|
 >{\centering\arraybackslash}p{4.0cm}|
 >{\centering\arraybackslash}p{1.5cm}|
 >{\centering\arraybackslash}p{2.0cm}|
 >{\centering\arraybackslash}p{2.0cm}|
 >{\centering\arraybackslash}p{1.5cm}|
 >{\centering\arraybackslash}p{2.0cm}|
 >{\centering\arraybackslash}p{2.0cm}|}
\hline
\hline
\parbox[c]{4.0cm}{\centering \textbf{NM}\\\textbf{Station}} &
\parbox[c]{1.5cm}{\centering \textbf{Days of}\\\textbf{Raw Data}} &
\parbox[c]{2.0cm}{\centering \textbf{Days of}\\\textbf{Removed Outliers}} &
\parbox[c]{2.0cm}{\centering \textbf{Removed Outliers}\\\textbf{Percentage}} &
\parbox[c]{1.5cm}{\centering \textbf{Days of}\\\textbf{Missing Data}} &
\parbox[c]{2.0cm}{\centering \textbf{Days of}\\\textbf{Imputed Data}} &
\parbox[c]{2.0cm}{\centering \textbf{Imputed Data}\\\textbf{Percentage}} \\
\hline
\hline
\textbf{AATB}  & 4,859 & 11  & 0.222\% & 103 & 114 & 2.297\%  \\
\textbf{APTY}  & 4,961 & 2   & 0.040\% & 1   & 3   & 0.060\%  \\
\textbf{FSMT}  & 4,948 & 5   & 0.101\% & 14  & 19  & 0.383\%  \\
\textbf{INVK}  & 4,916 & 1   & 0.020\% & 46  & 47  & 0.947\% \\
\textbf{JUNG}  & 4,951 & 5   & 0.101\% & 11  & 16  & 0.322\%  \\
\textbf{JUNG1} & 4,948 & 16  & 0.322\% & 14  & 30  & 0.605\%  \\
\textbf{LMKS}  & 4,954 & 23  & 0.464\% & 8   & 31  & 0.625\%  \\
\textbf{MXCO}  & 4,951 & 30  & 0.605\% & 11  & 41  & 0.826\%  \\
\textbf{NAIN}  & 4,942 & 1   & 0.020\% & 20  & 21  & 0.423\%  \\
\textbf{NEWK}  & 4,942 & 6   & 0.121\% & 20  & 26  & 0.524\%  \\
\textbf{OULU}  & 4,961  & 3   & 0.060\% & 1   & 4   & 0.081\%  \\
\textbf{PSNM}  & 4,909 & 0   & 0\%     & 53  & 53  & 1.068\%  \\
\textbf{PWNK}  & 4,939 & 1   & 0.020\% & 23  & 24  & 0.484\%  \\
\textbf{SOPB}  & 4,956 & 5   & 0.101\% & 6   & 11  & 0.222\%  \\
\textbf{SOPO}  & 4,952 & 3   & 0.060\% & 10  & 13  & 0.262\% \\
\textbf{TERA}  & 4,911 & 13  & 0.262\% & 51  & 64  & 1.290\%  \\
\textbf{THUL}  & 4,961 & 1   & 0.020\% & 1   & 2   & 0.040\%  \\
\textbf{YKTK}  & 4,932 & 28  & 0.564\% & 30  & 58  & 1.169\%  \\
\hline
\hline
\end{tabular}
\label{tab:Imputation1}
\end{table*}

\subsection{Accuracy of Neutron Monitor Missing Data Imputation}
\label{sec23}

We compare the imputation performance of various models within a unified framework on the NM dataset.
The performance comparison results are shown in Table \ref{table.5.1}.
In this experiment, the results indicate that both SAITS and iTransformer outperform TimesNet and Transformer in terms of MAE and RMSE.
While SAITS, Transformer, and iTransformer fundamentally rely on self-attention mechanisms to model temporal sequences, TimesNet adopts a different approach by transforming time series signals into two-dimensional time-frequency representations and employing a multi-scale modeling paradigm.
These findings suggest that self-attention-based methods are more adept at extracting intricate temporal and spectral features inherent in the NM count rates.
Moreover, when compared to the Transformer architecture, the enhanced iTransformer and SAITS exhibit greater efficiency in capturing subtle time-dependent patterns, thereby achieving improved performance.
Although iTransformer and SAITS show comparable performances in terms of key metrics (see Table \ref{table.5.1}), iTransformer demonstrates higher sensitivity to outliers during missing data reconstruction, which is undesirable.
Therefore, SAITS is adopted for missing data reconstruction in the subsequent experiments.
More details of the imputation results and visualizations are shown in Figure \ref{fig:imputation_result} in Appendix \ref{appendixB}.

After finalizing the imputation, we obtain continuous daily neutron count rates from 18 NM stations, covering the period from January 1, 2011, to August 1, 2024. The total number of days analyzed is 4,962.
A portion of this dataset is subsequently used to train the model that learns the relationship between ground-based neutron count rates and the space-based proton flux measurements.

\begin{table}[tp]
	\centering
	\caption{Performance comparison of four time series imputation methods on the NM dataset.}
    \setlength{\tabcolsep}{6mm}{
	\begin{tabular}{lcc}
		\hline
		\hline
		\textbf{Method} & \textbf{MAE} ($\downarrow$) & \textbf{RMSE} ($\downarrow$) \\
		\hline
		\hline
		TimesNet & 0.0274 & 0.0369 \\
        Transformer  & 0.0235 & 0.0326 \\
        iTransformer & 0.0126 & 0.0194 \\
        SAITS & \textbf{0.0121} & \textbf{0.0188} \\
		\hline
		\hline
	\end{tabular}
    }
	\label{table.5.1}%
\end{table}

The imputation of ground-based missing values to create continuous datasets is a critical step in periodic analysis.
This is because periodicity analysis relies on continuous data to accurately identify recurring patterns or cycles within a time series.

For a long time, the limitations of continuous data availability have led most studies to focus on analyzing data segments of shorter periods, which often makes it difficult to capture complete long-term trends or cyclical changes. 
Only a few studies have researched long-term variations.
For example, \citet{2019SoPh..294...15T} analyzed five solar cycles (1965-2018) using spectral methods, confirming known periodicities (11-year and 27-day) while discovering new patterns in cosmic rays and solar parameters (10-month and 3-year cycles).


Our study extends the continuous dataset to August 2024 by addressing the gap in recent measurements. This extension allows us to apply periodic analysis techniques to the latest observation, especially for in-depth research on the special phenomenon of polar magnetic field reversal in Solar Cycle 24. Filling the gaps in ground-based data not only eliminates the data breakpoints, but also creates conditions for applying more complex analysis methods, thereby revealing extensive periodic patterns that may not be observable in short-term datasets.

In terms of model training, continuous data is crucial for studying the relationship between the measurements of ground-based neutron monitors and AMS proton flux measurements. The lack of continuous data can make it difficult to evaluate time-dependence and accurately simulate the temporal correlation between these two types of measurements. The SAITS algorithm performs excellently in dealing with this kind of time series interpolation problem, and the supplemental neutron monitor count rates it generates significantly extends the available daily resolution datasets, providing valuable long-term observation resources for the research team.




\section{Details of Neutron Monitor Data Preprocessing}
\label{appendixA}

To ensure reliable and physically meaningful NM count rates for subsequent proton flux computation, we implement a comprehensive pre-processing pipeline consisting of multiple steps. These steps include obtaining high-resolution corrected NM count rates, applying robust statistical outlier detection using Interquartile Range (IQR), computing daily-averaged count rates from the dataset with filtering, and finally performing cross-checking of physically significant events across multiple NM stations. Each step is detailed below.

The NMDB provides access to cosmic-ray count rate records from 58 NM stations worldwide.
In this study, we adopt the corrected measurements, which account for atmospheric pressure variations and site-specific instrument configurations.
To obtain daily NM count rates suitable for model computation, we first extract 10-minute records and filter out outliers.
The traditional \( 3\sigma \) outlier removal method relies on the assumption that the data follows a normal distribution. However, the NM count rate does not exhibit a normal distribution. Therefore, we choose IQR which does not rely on any distributional assumptions over the \( 3\sigma \) method.

Specifically, we compute the first quartile $Q_1$ and the third quartile $Q_3$ of the count rates, and define $\text{IQR} = Q_3 - Q_1$.
Values outside the range $[Q_1 - 3 \cdot \text{IQR},\; Q_3 + 3 \cdot \text{IQR}]$ are considered as outliers and removed. We adopt a conservative $3 \cdot \text{IQR}$ threshold to minimize exclusion of real physical signals, while the traditional $1.5 \cdot \text{IQR}$ threshold is so tight that it removes real physical phenomena.

After removing outliers from 10-minute records, we compute daily-average count rates.
The same IQR-based procedure is then reapplied at the daily resolution to further remove any residual anomalies undetected at 10-minute timescales.


While IQR effectively removes outliers, it may unexpectedly exclude data points associated with real physical phenomena.
Solar-induced anomalies have been observed to occur concurrently at NM stations with similar geomagnetic cutoff rigidities. This phenomenon has been attributed to global cosmic‐ray modulation by solar disturbances, as documented in prior studies \citep{2015JGRA..120.5694K, 2003AdSpR..32..103S}.
To address this concern, we adopt a cross-station validation strategy. Specifically, if an outlier identified at one station coincides in time with similar anomalies at another stations with close geomagnetic cutoff rigidity, we consider it to be a  solar-related event and retain it in our dataset. 

In conclusion, this two-stage strategy, statistical filtering followed by cross-checking of physical events, ensures that the final daily dataset is physical.

\section{Imputation Results of Different Neutron Monitor Stations}
\label{appendixB}

\begin{figure*}[tp]
    \centering
    \includegraphics[width=\linewidth]{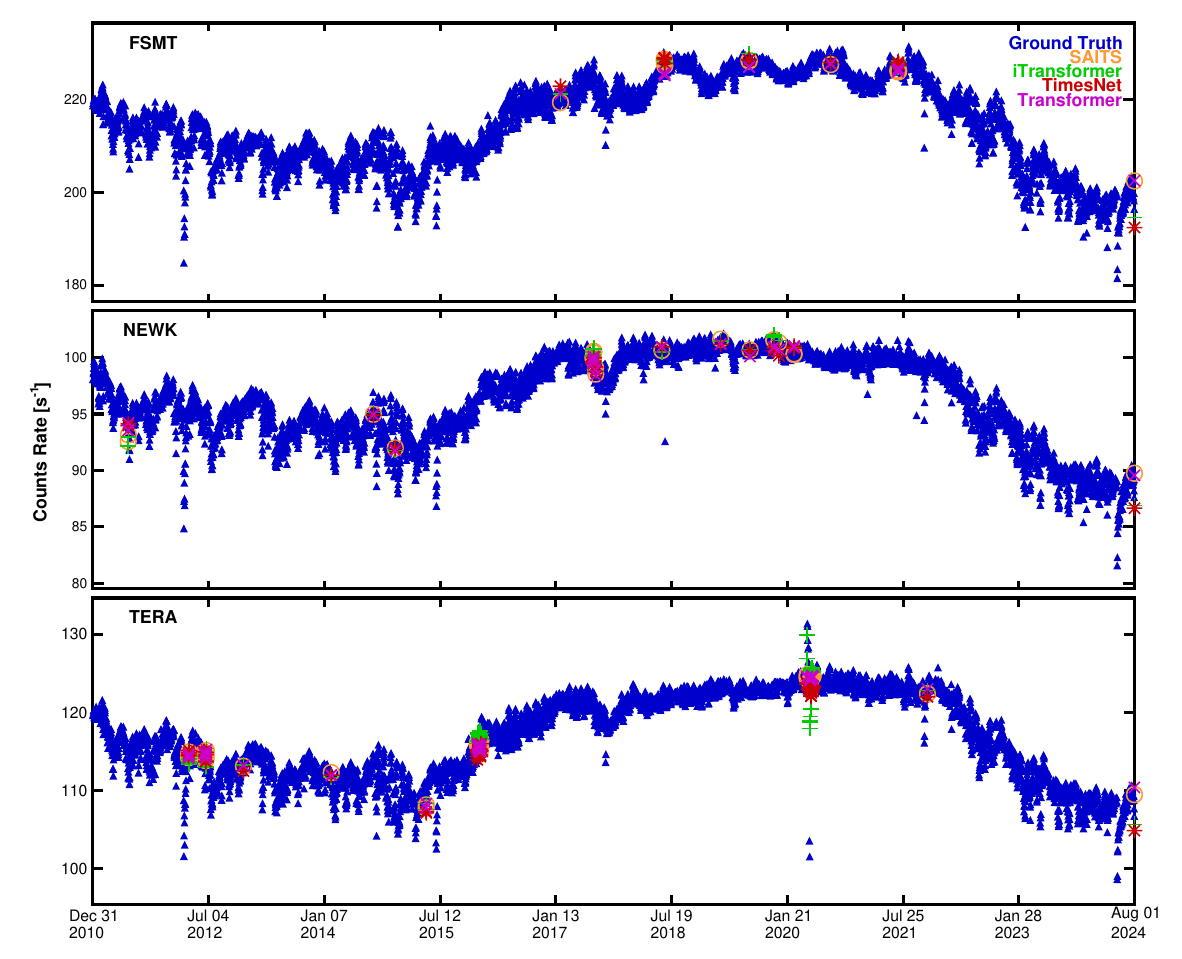}
    \caption{Comparative imputation performance of SAITS, TimesNet, Transformer and iTransformer models on temporal count rates from the \texttt{FSMT} (Top), \texttt{NEWK} (Middle) and \texttt{TERA} (Bottom) NM stations. The absolute count rate is plotted against time (x-axis), demonstrating the reconstruction capabilities of the four imputation methodologies.}

    \label{fig:imputation_result}
\end{figure*}

In Figure \ref{fig:imputation_result}, both SAITS and iTransformer successfully captured the underlying periodic patterns and temporal trends present in the raw NM count rates, with high fidelity to the original signal characteristics.
The imputed values, highlighted in yellow and green, closely align with the observed measurements, indicating robust performance in reconstructing missing temporal segments while preserving the intrinsic cyclical variations in cosmic ray intensity.

The imputation results in Figure \ref{fig:imputation_result} of \texttt{TERA} station further highlight the differences between SAITS and iTransformer.
iTransformer is more sensitive to outliers while SAITS is not.
Their differing feature extraction methods cause iTransformer to easily learn anomalous patterns from past time points, while SAITS demonstrates greater robustness.
This explains why the predicted points from SAITS are more concentrated, whereas the predicted points from iTransformer are more dispersed.
Although their performance metrics (mentioned in Table \ref{table.5.1}) are very close, as shown in Figure \ref{fig:imputation_result}, SAITS better meets the objectives of this work.

\bibliography{sample701}{}
\bibliographystyle{aasjournalv7}



\end{document}